\documentclass{llncs}
\usepackage{latexsym}

\def\gred{ {\bf g\_reduce}}

\def\grep{ {\bf g\_rep}}
\def\egnorm{ {\bf eg\_norm}}
\def\egred{ {\bf eg\_reduce}}
\def\egnrep{ {\bf eg\_normrep} }
\def\egrrep{ {\bf eg\_reducerep} }
\def\egrep{ {\bf eg\_rep}}

\title{An Extended Generalized Disjunctive Paraconsistent Data Model for Disjunctive Information}
\author{Haibin Wang, Hao Tian and Rajshekhar Sunderraman}
\institute{Department of Computer Science \\
Georgia State University \\ Atlanta, GA 30302 \\
email: \{hwang17,htian1\}@student.gsu.edu, raj@cs.gsu.edu}

\begin{document}
\maketitle

\bibliographystyle{splncs}

\begin{abstract}
This paper presents an extension of generalized disjunctive 
paraconsistent 
relational 
data  model in
which pure disjunctive positive and negative information 
as well as mixed disjunctive positive and negative information
can be represented
explicitly and manipulated. 
We consider explicit mixed disjunctive information in the context of 
disjunctive 
databases
as there is an interesting interplay between these two types
of information. {\em Extended generalized disjunctive paraconsistent 
relation}
is introduced as the main structure in this model. 
The relational algebra is appropriately
generalized to work on extended generalized disjunctive paraconsistent 
relations
and their correctness is established. 
\end{abstract}

\section{Introduction}

Relational data model was proposed by Ted Codd's pioneering paper~\cite{cdd70}.
Relational data model is value-oriented and has rich set of simple, but
powerful algebraic operators.
Moreover, a strong theoretical foundation for the model is provided  
by the classical first-order logic \cite{REIT84}.
This combination of a respectable theoretical platform, ease of
implementation and the practicality of the model resulted in its  
immediate success, and the model has enjoyed being used by many 
database management systems.

One limitation of the relational data model, however, is its lack of
applicability to nonclassical situations.
These are situations involving incomplete or even
inconsistent information.

Several types of incomplete information have been extensively 
studied in the past
such as {\em null} values \cite{cdd86,IL84,MAIE83}, 
{\em partial} values \cite{GRAN80}, {\em fuzzy} and 
{\em uncertain} values \cite{gln86,rjm88}, and 
{\em disjunctive} information \cite{LS88a,LS89a,snd93}.

In this paper, we present an extension of generalized disjunctive 
paraconsistent
data model\cite{WAN04}. 
Our model is capable of representing and
manipulating pure disjunctive positive facts and disjunctive negative
facts as well as mixed disjunctive positive and negative facts.
We introduce {\em extended generalized disjunctive paraconsistent relations}, 
which are 
the fundamental structures underlying our model.
These structures are extension of 
{\em generalized disjunctive paraconsistent relations}
which are capable of representing pure disjunctive positive
and negative facts. The motivation of this paper is that in the context of
disjunctive database, we should consider not only the pure disjunctive
information but also the mixed disjunctive information. The data model
represented in this paper can be applied to calculate the fixed point semantics
of extended disjunctive logic programs~\cite{MIN94} algebraically.  
An extended generalized disjunctive paraconsistent 
relation
essentially consists of three kinds of information:
positive tuple sets representing exclusive disjunctive positive facts 
(one of which
belongs to the relation), negative tuple sets representing exclusive disjunctive negated facts
(one of which does not belong to the relation) and mixed tuple sets 
representing exclusive disjunctive facts (one of which belongs to the relation
or one of which does not belong to the relation).
Extended generalized disjunctive paraconsistent relations are strictly more 
general than
generalized disjunctive paraconsistent relations in that for any generalized 
disjunctive paraconsistent relation,
there is an extended  generalized disjunctive paraconsistent relation with 
the same
information content, but not {\em vice versa}.
We extend the representation provided in~\cite{WAN04} by introducing
mixed disjunctive facts. There is an interesting interplay among these
three kinds of information.  
After introducing extended generalized disjunctive paraconsistent relations, 
we present
operators to remove redundancies and inconsistencies.
We also extend the standard relational algebra to operate on
extended generalized disjunctive paraconsistent relations. The information 
content of 
extended generalized disjunctive paraconsistent relations is characterized 
in terms of 
generalized disjucntive paraconsistent relations which we briefly present in 
the next 
section.

The rest of the paper is organized as follows.
Section~\ref{section2} provides the brief overview of generalized
disjunctive paraconsistent relations and the associated algebraic operators.
Section~\ref{section3} introduces extended generalized disjunctive 
paraconsistent relations as structures that are capable of representing
mixed disjunctive facts.
Section~\ref{section4} presents the notion of {\em precise generalization} of
algebraic operators and defines precise generalizations of several useful
algebraic operators. These operators can be used for specifying queries for
database systems built on such extended generalized disjunctive paraconsistent
relations. Finally, Section~\ref{section5} contains some concluding remarks
and directions for future work.

\section{Generalized Disjunctive Paraconsistent Relations}
\label{section2}

In this section, we present a brief overview of definition of generalized 
disjunctive 
paraconsistent
relations.
% and the algebraic operations on them. 
For a more detailed 
description, refer to \cite{WAN04}.

Let a {\em relation scheme} (or just {\em scheme})
$\Sigma$ be a finite set of {\em attribute names},
where for any attribute name
$A \in \Sigma$, $\mbox{\em dom}(A)$ is a non-empty
{\em domain} of values for $A$.
A {\em tuple} on $\Sigma$ is any map $t:\Sigma \rightarrow \cup_{A  
\in \Sigma}
\mbox{\em dom}(A)$, such that $t(A) \in \mbox{\em dom}(A)$, for each  
$A \in
\Sigma$.
Let $\tau(\Sigma)$ denote the set of all tuples on $\Sigma$.

\begin{definition}
\emph{A} paraconsistent relation \emph{on scheme} $\Sigma$ \emph{is a pair}
$R = \langle R^+, R^- \rangle$, \emph{where} $R^+$ \emph{and} $R^-$
\emph{are any subsets of} $\tau(\Sigma)$.
\emph{We let} ${\cal P}(\Sigma)$
\emph{be the set of all paraconsistent relations on}
$\Sigma$. \hfill{\space} $\Box$
\end{definition}

\begin{definition}
\emph{A} disjunctive paraconsistent relation, $R$, 
\emph{over the scheme}
$\Sigma$ \emph{consists of two components} $<R^{+}, R^{-}>$
\emph{where} $R^{+} \subseteq 2^{\tau(\Sigma)}$ \emph{and}
$ R^{-} \subseteq \tau(\Sigma)$.
$R^{+}$, \emph{the} positive \emph{component, is a set of tuple sets}.
\emph{Each tuple set in this component represents a disjunctive} 
\emph{positive fact. In the case where the} 
\emph{tuple set is a singleton, we have a definite positive fact.} 
$R^{-}$, \emph{the} negative \emph{component consists} 
\emph{of tuples that we refer to as definite negative tuples.} 
\emph{Let} ${\cal D}(\Sigma)$ \emph{represent all disjunctive paraconsistent}
\emph{relations over the scheme} $\Sigma$.
\hfill{\space} $\Box$
\end{definition}

\begin{definition}
\emph{A} generalized disjunctive paraconsistent relation, $R$, \emph{over the sc
heme}
$\Sigma$ \emph{consists of two components} $\langle R^+,R^- \rangle$ \emph{where
}
$R^+ \subseteq 2^{\tau(\Sigma)}$ \emph{and} $R^- \subset 2^{\tau(\Sigma)}$. $R^+
$,
\emph{the} positive \emph{component, is a set of tuple sets. Each tuple set in t
his
component represens a disjunctive positive fact. In the case where the tuple
set is a singleton, we have a definite positive fact}. $R^-$, \emph{the}
negative
\emph{component consists of a set of tuple sets. Each tuple set in this componen
t
represents a disjunctive negative fact. In the case where the tuple set is a
singleton, we have a definite negated fact. Let} ${\cal {GD}}(\Sigma)$ \emph{rep
resent
all generalized disjunctive paraconsistent relatios over the scheme} $\Sigma$.
\hfill{\space} $\Box$
\end{definition}

%\begin{definition}
%\emph{Let} $R$ \emph{be a generalized disjunctive paraconsistent relation over}
%$\Sigma$.
%$R^+ = \{w_1, w_2, \cdots, w_n \}$ \emph{and} $R^- = \{u_1, u_2, \cdots, u_m \}$
%.
%\emph{Then}, \\
%$\gnorm(R)^{+} = R^+ - \\ \hspace*{0.5in}
%\{w | w \in R^+ \wedge w \subseteq \cup u_i \wedge 1 \leq i \leq m \rightarrow u
%_i \in R^- \wedge |u_i| = 1\} - \\ \hspace*{0.5in}
%\{w_i | 1 \leq i \leq n \rightarrow w_i \in R^+ \wedge |w_i| = 1 \wedge (\exists
% u)(u \in R^- \wedge u \subseteq \cup w_i \wedge w_i \subseteq u)\}$ \\
%$\gnorm(R)^{-} = R^- - \\ \hspace*{0.5in}
%\{u | u \in R^- \wedge u \subseteq \cup w_i \wedge 1 \leq i \leq n \rightarrow
%w_i \in R^+ \wedge |w_i| = 1\} - \\ \hspace*{0.5in}
%\{u_i | 1 \leq i \leq m \rightarrow u_i \in R^- \wedge |u_i| = 1 \wedge (\exists
% w)(w \in R^+
%\wedge w \subseteq \cup u_i \wedge u_i \subseteq w)\}$
%\hfill{\space} $\Box$
%\end{definition}

A generalized disjunctive paraconsistent relation is called {\em normalized} if
it does not contain any inconsistencies. We let ${\cal {GN}}(\Sigma)$ denote the set
of all normalized generalized disjunctive paraconsistent relations over scheme
$\Sigma$.

\section{Extended Generalized Disjunctive Paraconsistent Relations}
\label{section3}

In this section, we present the main structure underling our model, the
{\em extended generalized disjunctive paraconsistent relations}. We identify 
several types of redundancies and inconsistencies that may appear and provide 
operators to remove them. Finally, we present the information content of 
extended generalized paraconsistent relations.

\begin{definition}
\emph{An}extended generalized disjunctive paraconsistent relation, $R$, 
\emph{over the scheme}
$\Sigma$ \emph{consists of three  components} $\langle R^+,R_M,R^- \rangle$ 
\emph{where} 
$R^+ \subseteq 2^{\tau(\Sigma)}$, \emph{each element} $r_M$ \emph{of} 
$R_M$ \emph{consists of two parts} $r_{M}^+ \in 2^{\tau(\Sigma)}$ \emph{and}
$r_{M}^- \in 2^{\tau(\Sigma)}$,
 \emph{and} $R^- \subset 2^{\tau(\Sigma)}$. $R^+$,
\emph{the} positive \emph{component, is a set of tuple sets. Each tuple set in this
component represens a disjunctive positive fact. In the case where the tuple
set is a singleton, we have a definite positive fact}. 
$R_M$, \emph{the} mixed \emph{component, is a set of pair tuple sets. The first
tuple set represents a disjunctive positive facts. The second tuple set represents
a disjunctive negated facts. And the relationship between these two tuple sets
is disjunctive.} 
$R^-$, \emph{the} 
negative
\emph{component consists of a set of tuple sets. Each tuple set in this component
represents a disjunctive negative fact. In the case where the tuple set is a
singleton, we have a definite negated fact. Let} ${\cal {EGD}}(\Sigma)$ 
\emph{represent
all extended generalized disjunctive paraconsistent relations over the scheme} 
$\Sigma$.
\hfill{\space} $\Box$
\end{definition}

%\begin{example}
%Consider the following generalized disjunctive paraconsistent relation:\\
%$supply^{+} = \{\{<s1,p1> \},\{<s2,p1>, <s2,p2> \},\{<s3,p3>, <s3,p4> \} \}$ \\
%$supply^{-} = \{\{<s1,p2> \}, \{<s1,p3>\}, \{<s2,p3>, <s2,p4> \}\}$.
%The {\em positive component} corresponds to the statement $s1$ supplies 
%$p1$, $s2$ supplies $p1$ or $p2$, and $s3$ supplies $p3$ or $p4$ and the 
%{\em negative component} corresponds to $s1$ does not supply $p2$ and $s1$ 
%does not supply $p3$ and $s2$ does not supply $p3$ or $s2$ does not 
%supply $p4$. It should be noted that the status of tuples that do not appear 
%anywhere in the generalized disjunctive paraconsistent relation, such as 
%$(s3,p2)$, is unknown.
%\hfill{\space} $\Box$
%\end{example}

Inconsistences can be present in an extended genearlaized disjunctive 
paraconsistent
relation in three situations. First, if all the tuples of a tuple set 
of the posistive component are also present in the union of the singleton tuple
set of the negative component. 
%In such a case, the tuple set states that at 
%least one of the tuples in the tuple set must be in the relation whereas the 
%negative component states that all the tuples in the tuple set must not be in
%the relation. 
We deal with this inconsistency by removing both the positive
tuple set and all its corresponding singleton tuple sets from the negative
component. Second, if all the tuples of a tuple set of the negative
component are also present in the union of the singleton tuple set of the 
positive component. 
%In such a case, the tuple set states that at least one of
%the tuples in the tuple set must not be in the relation whereas the positive
%component states that all the tuples in the tuple set must be in the relation.
We deal with this inconsistency by removing both the negative tuple set and
all its corresponding singleton tuple sets from the positive component. 
Third, if all the tuples of the first tuple set of the pair tuple sets of
mixed component are also present in the union of the singleton tuple set
of the negative component and all the tuples of the second tuple set of the
pair tuple sets of mixed component are also present in the union of the
singleton tuple set of the positive component. We deal with this inconsistency
by removing the pair tuple sets and its corresponding singleton tuple sets
from the negative component and its corresponding singleton tuple sets from
the positive component. 
This
is done by the $\egnorm$ operator defined as follows:

\begin{definition} 
\emph{Let} $R$ \emph{be an extended generalized disjunctive paraconsistent relation over} 
$\Sigma$. 
$R^+ = \{w_1, w_2, \cdots, w_n \}$, $R_M = \{<v_1,x_1>, \cdots, <v_k,x_k>\}$ 
\emph{and} $R^- = \{u_1, u_2, \cdots, u_m \}$.
\emph{Then}, \\
$\egnorm(R)^{+} = R^+ - \\  
\{w | w \in R^+ \wedge w \subseteq \cup u_i \wedge 1 \leq i \leq m \rightarrow u_i \in R^- \wedge |u_i| = 1\} - \\ 
\{w_i | 1 \leq i \leq n \rightarrow w_i \in R^+ \wedge |w_i| = 1 \wedge (\exists u)(u \in R^- \wedge u \subseteq \cup w_i \wedge w_i \subseteq u)\} - \\ 
\{w_i | 1 \leq i \leq n \rightarrow w_i \in R^+ \wedge |w_i| = 1 \wedge (\exists \langle v,x \rangle)(\langle v,x \rangle \in R_M \wedge x \subseteq \cup w_i \wedge w_i \subseteq x \wedge v \subseteq \cup u_j \wedge u_j \in R^- \wedge |u_j| = 1)\}$ \\
$\egnorm(R)_M = R_M - \\ 
\{\langle v,x \rangle | \langle v,x \rangle \in R_M \wedge v \subseteq \cup u_i \wedge 1 \leq i \leq m \rightarrow u_i \in R^- \wedge |u_i| = 1 \wedge x \subseteq \cup w_j \wedge 1 \leq j \leq n \rightarrow w_j \in R^+ \wedge |w_j| = 1\}$ \\ 
$\egnorm(R)^{-} = R^- - \\
\{u | u \in R^- \wedge u \subseteq \cup w_i \wedge 1 \leq i \leq n \rightarrow  
w_i \in R^+ \wedge |w_i| = 1\} - \\ 
\{u_i | 1 \leq i \leq m \rightarrow u_i \in R^- \wedge |u_i| = 1 \wedge (\exists w)(w \in R^+ 
\wedge w \subseteq \cup u_i \wedge u_i \subseteq w)\} - \\
\{u_i | 1 \leq i \leq n \rightarrow u_i \in R^- \wedge |u_i| = 1 \wedge (\exists \langle v,x \rangle)(\langle v,x \rangle \in R_M \wedge v 
\subseteq \cup u_i \wedge u_i \subseteq v \wedge x \subseteq \cup w_j \wedge w_j \in R^+ \wedge |w_j| = 1)\}$  

\hfill{\space} $\Box$
\end{definition}

An extended generalized disjunctive paraconsistent relation is called {\em normalized} if
it does not contain any inconsistencies. We let ${\cal {EGN}}(\Sigma)$ denote the set
of all normalized extended generalized disjunctive paraconsistent relations over scheme
$\Sigma$.
We now identify the following eight types of redundancies in a normalized extended 
generalized disjunctive paraconsistent relation $R$: \\

\vspace{-.1in}

\begin{enumerate}
\item \underline{$w_1 \in R^+$, $w_2 \in R^+$, and $w_1 \subset w_2$.}
In this case, $w_1$ subsumes $w_2$. To eliminate this redundancy, we delete
$w_2$ from $R^+$.

\item \underline{$u_1 \in R^-$, $u_2 \in R^-$, and $u_1 \subset u_2$.}
In this case, $u_1$ subsumes $u_2$. To eliminate this redundancy, we delete
$u_2$ from $R^-$.

\item \underline{$1 \leq i \leq n$, $w_i \in R^+$, $|w_i| = 1$, $u \in R^-$, and $\cup w_i \subset u$.} This redundancy is eliminated by deleting the tuple set
$u$ from $R^-$ and adding the tuple set $u - \cup w_i$ to $R^-$. Since we are
dealing with normalized generalized disjunctive paraconsistent relations,
$u - \cup w_i$ cannot be empty.

\item \underline{$1 \leq i \leq m$, $u_i \in R^-$, $|u_i| = 1$, $w \in R^+$, and $\cup u_i \subset w$.} This redundancy is eliminated by deleting the tuple set
$w$ from $R^+$ and adding the tuple set $w - \cup u_i$ to $R^+$. Since we are
dealing with normalized generalized disjunctive paraconsistent relations,
$w - \cup u_i$ cannot be empty. 

\item \underline{$\langle v_1,x_1 \rangle \in R_M$, $\langle v_2,x_2 \rangle \in R_M$, $v_1 \subset v_2$ and $x_1 \subset x_2$.}
In this case, $\langle v_1,x_1 \rangle$ subsumes $\langle v_2,x_2 \rangle$. To eliminate this redundancy, we delete $\langle v_2,x_2 \rangle$
from $R_M$.

\item \underline{$\langle v,x \rangle \in R_M$, $w \in R^+$, and $w \subseteq v$.}
In this case, $w$ subsumes  $v \vee x$. To eliminate this redundancy, we delete $\langle v,x \rangle$ from $R_M$.

\item \underline{$\langle v,x \rangle \in R_M$, $u \in R^-$, and $u \subseteq x$.}
In this case, $u$ subsumes $v \vee x$. To eliminate this redundancy, we delete $\langle v,x \rangle$ from $R_M$.

\item \underline{$1 \leq i \leq n$, $w_i \in R^+$, $|w_i| = 1$, $1 \leq j \leq m$, $u_j \in R^-$, $|u_j| = 1$,  
$\langle v,x \rangle \in R_M$,} \\
\underline{$\cup u_j \cap v \neq \emptyset$ or $\cup w_i \cap x \neq \emptyset$.} 
This redundancy is eliminated by deleteing the pair tuple sets
$\langle v,x \rangle$ from $R_M$. And addes the pair tuple sets $\langle v - \cup u_j, x - \cup w_i \rangle$ to $R_M$ if $v - \cup u_j$ and $x - \cup w_i$
are not empty. If $v - \cup u_j$ is empty then addes the tuple set $x - \cup w_i$ to $R^-$. If $x - \cup w_i$ is empty then adds the tuple
set $v - \cup u_j$ to $R^+$. Since we are dealing with normalized generalized disjunctive paraconsistent relations, $x - \cup w_i$ and
$v - \cup u_j$ cannot be both empty. 
\end{enumerate}
We now introduce an operator called $\egred$ to take care of redundancies.

\begin{definition}
\emph{Let} $R$ \emph{be a normalized extended generalized disjunctive paraconsistent
relation. Then,} \\
$\egred(R)^+ = \{w' | (\exists w) (w \in R^+ \wedge w' = w - U \wedge$ 
$\neg (\exists w_1) (w_1 \in R^+ \wedge (w_1 - U) \subset w')) \} \cup \{w' | (\exists \langle v,x \rangle)(\langle v,x \rangle \in R_M \wedge w' = v - U \wedge x - W = \emptyset)\}$ \\
$\egred(R)_M = \{\langle v',x' \rangle | (\exists \langle v,x \rangle)(\langle v,x \rangle \in R_M \wedge \neg (\exists w)(w \in R^+ \wedge w \subseteq v) \wedge \neg (\exists u)(u \in R^- \wedge u \subseteq x) \wedge v' = v - U \wedge x' = x - W \wedge v - U \neq \emptyset \wedge x - W \neq \emptyset \wedge \neg (\exists \langle v_1,x_1 \rangle)(\langle v_1,x_1 \rangle \in R_M \wedge (v_1 - U) \subset v' \wedge (x_1 - W) \subset x')\}$ \\
$\egred(R)^- = \{u' | (\exists u) (u \in R^- \wedge u' = u - W \wedge$ 
$\neg (\exists u_1) (u_1 \in R^- \wedge (u_1 -W) \subset u')) \} \cup \{u' | (\exists \langle v,x \rangle)(\langle v,x \rangle \in R_M \wedge u' = x - W \wedge v - U = \emptyset)\}$ \\ 
\emph{where}, $U = \{u_i | u_i \in R^- \wedge |u_i| = 1\}$ \emph{and} $W = \{w_i | w_i \in R^+ \wedge |w_i| = 1\}$.
\hfill{\space} $\Box$ 
\end{definition}

%\begin{example}
%Consider the following generalized disjunctive paraconsistent relation:
%$R^+ = \{ \{<a>\}, \{<b>,<c>\}, \{<c>,<d>\}, \{<a>,<e>\},\{<f>,<g>\} \}$ \\
%and $R^- = \{ \{<b>\}, \{<c>,<e>\}, \{<i>\}, \{<d>,<e>,<f>\} \}$.
%The disjunctive tuple $\{<a>,<e>\}$ is subsumed by $\{<a>\}$ and hence removed.
%In the disjunctive tuple set $\{<b>,<c>\}$, $<b>$ is redundant due to the 
%presence of the negative singleton tuple set $\{<b>\}$ resulting in the positive tuple
%$\{<c>\}$ which in turn subsumes $\{<c>,<d>\}$ and makes $\{<c>,<e>\}$
%redundant and resulting in $\{<e>\}$ which subsumes the $\{<d>,<e>,<f>\}$.
%The reduced generalized disjunctive paraconsistent relation is:
%$\gred(R)^+ = \{ \{<a>\}, \{<c>\}, \{<f>,<g>\} \}$ and
%$\gred(R)^- = \{ \{<b>\}, \{<e>\}, \{<i>\} \}$
%\hfill{\space} $\Box$
%\end{example}

The information content of an extended generalized disjunctive paraconsistent 
relation
can be defined to be a collection of generalized disjunctive paraconsistent 
relations.
The different possible generalized disjunctive paraconsistent relations are 
constructed
by selecting one of the tuple sets within a pair tuple sets for each pair tuple 
sets
in the mixed component. In doing so, we may end up with non-minimal generalized
disjunctive paraconsistent relations or even with inconsistent generalized 
disjunctive
paraconsistent relations. These would have to be removed in order to obtain
the exact information content of extended generalized disjunctive 
paraconsistent
relations. The formal definitions follow:

\begin{definition}
\emph{Let} $U \subseteq {\cal {GD}}(\Sigma)$. \emph{Then},
$\egnrep_\Sigma(U) = \{R | R \in U \wedge \neg (\exists w)(w \in R^+ \wedge w \subseteq \cup u_i \wedge u_i \in R^- \wedge |u_i| = 1) \wedge \neg (\exists u)(u \in R^- \wedge u \subseteq \cup w_i \wedge w_i \in R^+ \wedge |w_i| = 1)\}$
\hfill{\space} $\Box$
\end{definition}

The $\egnrep$ operator removes all inconsistent generalized disjunctive paraconsistent 
relations from its input.

\begin{definition}
\emph{Let} $U \subseteq {\cal {GD}}(\Sigma)$. \emph{Then},
$\egrrep_\Sigma(U) = \{R | R \in U \wedge \neg (\exists S)(S \in U \wedge R \neq S \wedge S^+ \subseteq R^+ \wedge S^- \subseteq R^-)\}$
\hfill{\space} $\Box$
\end{definition}

The $\egrrep$ operator keeps only the ``minimal'' generalized disjunctive paraconsistent relations and eliminates any 
generalized disjunctive paraconsistent relation that is ``subsumed'' by others.

\begin{definition}
\emph{The information content of extended generalized disjunctive paraconsistent 
relations is defined by the mapping} $\egrep_\Sigma ~:~ {\cal {EGN}}(\Sigma) \rightarrow {\cal {GD}}(\Sigma)$. 
\emph{Let} $R$ \emph{be a normalized extended generalized disjunctive
paraconsistent relation on scheme} $\Sigma$ with $R_M = \{\langle v_1,x_1 \rangle \ldots, \langle v_k,x_k \rangle\}$.
\emph{Let} $U = \{R^+ \cup V, R^- \cup X | V = \{v_i | 1 \leq i \leq k\} \wedge X = \{x_j | 1 \leq j \leq k\} \wedge i \neq j \wedge |V| + |X| = k\}$. \emph{Then},
$\egrep_\Sigma(R) = \egrrep_\Sigma(\egnrep_\Sigma(U))$
\hfill{\space} $\Box$ 
\end{definition}

Note that the information content is defined only for normalized extended 
generalized 
disjunctive paraconsistent relations.

%\begin{example}
%Consider the following generalized disjunctive paraconsistent relation on a
%single attribute scheme $\Sigma$:
%$R^+ = \{ \{<b>,<e>\},\{<c>,<d>\},\{<e>,<g>\} \} and R^- = \{ \{<b>\}, \{<c>,<e>\}, \{<c>, <d>,<g>\}\}$
%The process of selecting tuples from tuple sets produces the following 
%disjunctive paraconsistent relations:

%$U = \{ <\{\{ \{<b>,<e>\}, \{<c>,<d>\}, \{<e>,<g>\}\}, \{<b>,<c>\} \} >, 
%        <\{\{ \{<b>,<e>\}, \{<c>,<d>\}, \{<e>,<g>\}\}, \{<b>,<c>,<d>\} \} >,
%        <\{\{ \{<b>,<e>\}, \{<c>,<d>\}, \{<e>,<g>\}\}, \{<b>,<c>,<g>\} \} >, 
%        <\{\{ \{<b>,<e>\}, \{<c>,<d>\}, \{<e>,<g>\}\}, \{<b>,<e>,<c>\} \} >,
%        <\{\{ \{<b>,<e>\}, \{<c>,<d>\}, \{<e>,<g>\}\}, \{<b>,<e>,<d>\} \} >,
%        <\{\{ \{<b>,<e>\}, \{<c>,<d>\}, \{<e>,<g>\}\}, \{<b>,<e>,<d>\} \} > \}$.

%Normalizing the above set of disjunctive paraconsistent relations using \\
%$\gnrep$ gives us:
%$U' = \{<\{\{ \{<b>,<e>\}, \{<c>,<d>\}, \{<e>,<g>\}\}, \{<b>,<c>\} \} >,
%        <\{\{ \{<b>,<e>\}, \{<c>,<d>\}, \{<e>,<g>\}\}, \{<b>,<c>,<g>\} \} >\}$. 

%Finally, removing the non-minimal disjunctive paraconsistent relations using 
%the $\grrep$ operator, we get the information content $\grep_\Sigma(R)$ as 
%follows:
%$\grep_\Sigma(R) = \{<\{\{ \{<b>,<e>\}, \{<c>,<d>\}, \{<e>,<g>\}\}, \{<b>,<c>\} \}>\}$.
%\hfill{\space} $\Box$        
%\end{example}

The following important theorem states that information is neither lost nor 
gained by removing the redundancies in an extended generalized disjunctive 
paraconsistent
relations.

\begin{theorem}
Let $R$ be an extended generalized disjunctive paraconsistent relation on
scheme $\Sigma$. Then, \\
$\egrep_{\Sigma}(\egred(R)) = \egrep_{\Sigma}(R)$
\hfill{\space} $\Box$
\end{theorem}

\section{Generalized Relational Algebra}
\label{section4}

In this section, we first develop the notion of {\em precise generalizations}
of algebraic operators. This is an important property that must be satisfied
by any new operator defined for extended generalized disjunctive 
paraconsistent 
relations. Then, we present several algebraic operators on extended generalized
disjunctive paraconsistent relations that are precise generalizations of their
counterparts on generalized disjunctive paraconsistent relations.

%\subsection*{Precise Generalization of Operations}

%It is easily seen that generalized disjunctive paraconsistent relations are a 
%generalization of disjunctive paraconsistent relations, in that for each 
%disjunctive paraconsistent relation there is a generalized disjunctive
%paraconsistent relation with the same information content, but not 
%{\em vice versa}. It is thus natural to think of generalising the operations
%on disjunctive paraconsistent relations, such as union, join, projection etc.,
%to generalized disjunctive paraconsistent relations. However, any such 
%generalization should be intuitive with respect to the {\em belief system}
%model of generalized disjunctive paraconsistent relations. We now construct a
%framework for operators on both kinds of relations and introduce the notion of
%the precise generalization relationship among their operators.

An $n$-ary {\em operator on generalized disjunctive paraconsistent relations 
with 
signature} \\ $\langle \Sigma_1, \ldots, \Sigma_{n+1} \rangle$ is a function $\Theta~:~{\cal {GD}}(\Sigma_1) \times \cdots \times {\cal {GD}}(\Sigma_n) 
\rightarrow {\cal {GD}}(\Sigma_{n+1})$, where $\Sigma_1, \ldots, \Sigma_{n+1}$ are any schemes. Similarly, an $n$-ary
{\em operator on extended generalized disjunctive paraconsistent relations with signature} $\langle \Sigma_1, \ldots, \Sigma_{n+1} \rangle$ is a function: $\Psi: {\cal {EGD}}(\Sigma_1) \times \cdots \times {\cal {EGD}}(\Sigma_n) 
\rightarrow {\cal {EGD}}(\Sigma_{n+1})$.

We now need to extend operators on generalized disjunctive paraconsistent 
relations to 
sets of generalized disjunctive paraconsistent relations. 
For any operator $\Theta~:~{\cal 
{GD}}(\Sigma_1) \times \cdots \times {\cal {GD}}(\Sigma_n) \rightarrow 
{\cal {GD}}(\Sigma
_{n+1})$ on generalized disjunctive paraconsistent relations, we let ${\cal S}(\Theta)~:~ 2^{{\cal {GD}}(\Sigma_1)} \times \cdots \times 2^{{\cal {GD}}(\Sigma_n)} \rightarrow 2^{{\cal {GD}}(\Sigma_{n+1})}$ be a map on sets of generalized
disjunctive paraconsistent relations
defined as follows. For any sets $M_1, \ldots, M_n$ of generalized disjunctive 
paraconsistent
relations on schemes $\Sigma_1, \ldots, \Sigma_n$, respectively, \\

${\cal S}(\Theta)(M_1, \ldots, M_n) = \{\Theta(R_1, \ldots, R_n) | R_i \in M_i, \mbox{for all } i, 1 \leq i \leq n\}$. \\

In other words, ${\cal S}(\Theta)(M_1, \ldots, M_n)$ is the set of $\Theta$-images of all tuples in the cartesian product $M_1 \times \cdots \times M_n$. We
are now ready to lead up to the notion of precise operator generalization.

\begin{definition}
\emph{An operator} $\Psi$ \emph{on extended generalized disjunctive 
paraconsistent
relations with signature} $\langle \Sigma_1, \ldots, \Sigma_{n+1} \rangle$ 
\emph{is} consistency preserving \emph{if for any normalized extended 
generalized 
disjunctive relations} $R_1, \ldots, R_n$ \emph{on schemes} $\Sigma_1, \ldots,
\Sigma_n$, \emph{respectively}, $\Psi(R_1, \ldots, R_n)$ \emph{is also 
normalized}.
\hfill{\space} $\Box$
\end{definition}

\begin{definition}
\emph{A consistency preserving operator} $\Psi$ \emph{on extended generalized 
disjunctive paraconsistent relations with signature} $\langle \Sigma_1, \ldots, \Sigma_{n+1} \rangle$ 
\emph{is a} precise generalization \emph{of an operator} $\Theta$ \emph{on generalized disjunctive paraconsistent relations with the same signature, 
if for any normalized extended 
generalized disjunctive paraconsistent relations} $R_1, \ldots, R_n$ 
\emph{on schemes} $\Sigma_1, \ldots, \Sigma_n$, \emph{we have} \\ 
$\egrep_{\Sigma_{n+1}}(\Psi(R_1, \ldots, R_n)) = {\cal S}(\Theta)(\egrep_{\Sigma_1}(R_1), \ldots, \grep_{\Sigma_n}(R_n))$.

\hfill{\space} $\Box$
\end{definition}

We now present precise generalizations for the usual relation operators, such as union, join, projection. 
To reflect generalization, a dot is placed over an
ordinary operator. For example, $\Join$ denotes the natural join among ordinary
relations, 
$\overline{\Join}$ denotes natural join on generalized disjunctive 
paraconsistent relations and $\dot{\Join}$ denotes natural join on extended 
generalized disjunctive paraconsistent relations.

\begin{definition}
\emph{Let} $R$ \emph{and} $S$ \emph{be two normalized extended 
generalized disjunctive
paraconsistent relations on scheme} $\Sigma$ \emph{with} 
$R_M = \{\langle p_1,n_1 \rangle, \ldots, \langle p_k,n_k \rangle\}$
\emph{and} 
$S_M = \{\langle u_1,v_1 \rangle, \ldots, \langle u_m,v_m \rangle\}$. 
\emph{Then}, 
$R \dot{\cup} S$
is an extended generalized disjunctive paraconsistent relation over 
scheme $\Sigma$ 
given by $R \dot{\cup} S = \egred(T)$, \emph{where} $T$ \emph{is defined
as follows}. \emph{Let} $E = \{\langle \egred(R)^+ \cup P,\egred(R)^- \cup N 
\rangle | P = \{p_i | (\forall i) \langle p_i,n_i \rangle \in \egred(R)_M \} 
\wedge N = \{n_j | (\forall j) \langle p_j,n_j \rangle \in$ \\ 
$\egred(R)_M \} 
\wedge i \neq j \wedge |P| + |N| = |\egred(R)_M|
 \}$ \emph{and} \\ 
$F = \{\langle \egred(S)^+ \cup U,\egred(S)^- \cup V \rangle | U = \{u_i | (\forall i) \langle u_i,v_i \rangle \in$ \\
$\egred(S)_M \} \wedge V = \{v_j | (\forall j) \langle u_j,v_j \rangle \in \egred(S)_M \wedge i \neq j \wedge |U| + |V| = |\egred(S)_M| \}$. \emph{Let the normalized elements of} $E$ 
\emph{be} $E_1, \ldots, E_e$ \emph{and those of} $F$ 
\emph{be} $F_1, \ldots, F_f$
\emph{and let} $A_{ij} = E_i \overline{\cup} F_j$, 
\emph{for} $1 \leq i \leq e$ \emph{and}
$1 \leq j \leq f$. 
\emph{Let} $A_1, \ldots, A_g$ \emph{be the distinct} $A_{ij}$s. 
\emph{Then}, \\ 
$T^+ = \{w | (\exists t_1) \cdots (\exists t_g)(t_1 \in A_{1}^+ \wedge \cdots \wedge t_g \in A_{g}^+ \wedge w = \{t_1, \ldots, t_g\})\}$ \\ 
$T_M = \{\langle P,N \rangle | P = \{p_i | (\forall i)p_i \in A_{i}^+\} \wedge N = \{n_j | (\forall j)n_j \in A_{j}^-\} \wedge i \neq j \wedge |p| \neq 0 \wedge |N| \neq 0 \wedge |P| + |N| = g\}$ \\
$T^- = \{u|(\exists t_1) \cdots (\exists t_g)(t_1 \in A_{1}^- \wedge \cdots \wedge t_g \in A_{g}^- \wedge u = \{t_1, \ldots, t_g\})\}$. \\ 
\emph{and} $R \dot{\cap} S$ is an extended 
generalized disjunctive paraconsistent relation over scheme $\Sigma$ given by
$R \dot{\cap} S = \egred(T)$, \emph{where} $T$ \emph{is defined
as follows}. 
\emph{Let} $B_{ij} = E_i \overline{\cap} F_j$, 
\emph{for} $1 \leq i \leq e$ \emph{and}
$1 \leq j \leq f$. \emph{Let} $B_1, \ldots, B_g$ \emph{be the distinct} \\ 
$B_{ij}$s
\emph{Then}, \\
$T^+ = \{w | (\exists t_1) \cdots (\exists t_g)(t_1 \in B_{1}^+ \wedge \cdots 
\wedge t_g \in B_{g}^+ \wedge w = \{t_1, \ldots, t_g\})\}$ \\
$T_M = \{\langle P,N \rangle | P = \{p_i | (\forall i)p_i \in B_{i}^+\} \wedge 
 N = \{n_j | (\forall j)n_j \in B_{j}^-\} \wedge i \neq j \wedge |p| \neq 0 
 \wedge |N| \neq 0 \wedge |P| + |N| = g\}$ \\
$T^- = \{u|(\exists t_1) \cdots (\exists t_g)(t_1 \in B_{1}^- \wedge \cdots 
 \wedge t_g \in B_{g}^- \wedge u = \{t_1, \ldots, t_g\})\}$. 

\hfill{\space} $\Box$
\end{definition}

The following theorem establishes the {\em precise generalization} property 
for union and intersection:

\begin{theorem}
Let $R$ and $S$ be two normalized extended generalized disjunctive
paraconsistent relations on scheme $\Sigma$. Then,

\vspace{-.1in}

\begin{enumerate}
\item $\egrep_{\Sigma}(R \dot{\cup} S) = \egrep_{\Sigma}(R) {\cal S}(\overline{\cup})\egrep_{\Sigma}(S)$.
\item $\egrep_{\Sigma}(R \dot{\cap} S) = \egrep_{\Sigma}(R) {\cal S}
(\overline{\cup})\egrep_{\Sigma}(S)$.
\hfill{\space} $\Box$
\end{enumerate}
\end{theorem}

\begin{definition}
\emph{Let} $R$ \emph{be normalized extended generalized disjunctive 
paraconsistent
relation on scheme} $\Sigma$. \emph{Then}, $\dot{-}R$ \emph{is an extended
generalized
disjunctive paraconsistent relation over scheme} $\Sigma$ \emph{given by}
 $(\dot{-}
R)^+ = \egred(R)^-$, $(\dot{-}R)_M = \{\langle p_i,n_i \rangle |(\forall i)
\langle n_i,p_i \rangle \in \egred(R)_M\}$ \emph{and} $(\dot{-}R)^- =  
\egred(R)^+$.

\hfill{\space} $\Box$
\end{definition}

\begin{definition}
\emph{Let} $R$ \emph{be a normalized extended generalized disjunctive 
paraconsistent 
relation on scheme} $\Sigma$ \emph{with}
$R_M = \{\langle p_1,n_1 \rangle, \ldots, \langle p_k,n_k \rangle\}$
, \emph{and let} $F$ \emph{be any logic formula 
involving attribute names in} $\Sigma$, \emph{constant symbols (denoting 
values in the attribute domains), equality symbol} $=$, \emph{negation symbol} 
$\neg$, \emph{and connectives} $\vee$ \emph{and} $\wedge$. \emph{Then, the}
selection of $R$ by $F$, denoted $\dot{\sigma}_F(R)$, \emph{is an extended 
generalized disjunctive paraconsistent relation on scheme} $\Sigma$,
\emph{given by} $\dot{\sigma}_F(R) = \egred(T)$, \emph{where} $T$ \emph{is defined as follows}. \\ 
\emph{Let} $E = \{\langle \egred(R)^+ \cup P,\egred(R)^- \cup N
\rangle | P = \{p_i | (\forall i) \langle p_i,n_i \rangle \in \egred(R)_M \}
\wedge N = \{n_j | (\forall j) \langle p_j,n_j \rangle \in$ \\
$\egred(R)_M \}
\wedge i \neq j \wedge |P| + |N| = |\egred(R)_M|
 \}$.  \\
\emph{Let the normalized
elements of} $E$
\emph{be} $E_1, \ldots, E_e$ \emph{and let} $A_{i} = \overline{\sigma}_F(E_i)$,
\emph{for} $1 \leq i \leq e$.  
\emph{Let} $A_1, \ldots, A_g$ \emph{be the distinct} $A_{i}$s.
\emph{Then}, \\
$T^+ = \{w | (\exists t_1) \cdots (\exists t_g)(t_1 \in A_{1}^+ \wedge \cdots 
\wedge t_g \in A_{g}^+ \wedge w = \{t_1, \ldots, t_g\})\}$ \\
$T_M = \{\langle P,N \rangle | P = \{p_i | (\forall i)p_i \in A_{i}^+\} 
\wedge N = \{n_j | (\forall j)n_j \in A_{j}^-\} \wedge i \neq j \wedge |p| 
\neq 0 \wedge
 |N| \neq 0 \wedge |P| + |N| = g\}$ \\
$T^- = \{u|(\exists t_1) \cdots (\exists t_g)(t_1 \in A_{1}^- \wedge \cdots 
\wedge t_g \in A_{g}^- \wedge u = \{t_1, \ldots, t_g\})\}$. 
\hfill{\space} $\Box$
\end{definition}

%A disjunctive tuple set is either selected as a whole or not at all. All the
%tuples within the tuple set must satisfy the selection criteria for the tuple
%set to be selected.

%If $\Sigma$ and $\Delta$ are relation schemes such that $\Sigma \subseteq \Delta
%$, then for any tuple $t \in \tau(\Sigma)$, we let $t^\Delta$ denote the set $\{
%t' \in \tau(\Delta) ~|~ t'(A) = t(A) \mbox{, for all $A \in \Sigma$}\}$ of all e
%xtensions of $t$. We extend this notion for any $T \subseteq \tau(\Sigma)$ by de
%fining $T^\Delta = \cup_{t \in T} ~ t^\Delta$.

\begin{definition}
\emph{Let} $R$ \emph{be a normalized extended generalized disjunctive 
paraconsistent 
relation on scheme} $\Sigma$ \emph{with} 
$R_M = \{\langle p_1,n_1 \rangle, \ldots, \langle p_k,n_k \rangle\}$,
\emph{and} $\Delta \subseteq \Sigma$. 
\emph{Then, the} projection of $R$ onto $\Delta$, denoted $\dot{\pi}_\Delta(R)$, \emph{is a generalized extended disjunctive paraconsistent relation on scheme}
$\Delta$, \emph{given by} $\dot{\pi}_\Delta(R) = \egred(T)$, \emph{where}
$T$ \emph{is defined as follows}. 
\emph{Let} $E = \{\langle \egred(R)^+ \cup P,\egred(R)^- \cup N
\rangle | P = \{p_i | (\forall i) \langle p_i,n_i \rangle \in \egred(R)_M \}
\wedge N = \{n_j | (\forall j) \langle p_j,n_j \rangle \in$ \\
$\egred(R)_M \}
\wedge i \neq j \wedge |P| + |N| = |\egred(R)_M|
 \}$.  \\
\emph{Let the normalized
elements of} $E$
\emph{be} $E_1, \ldots, E_e$ \emph{and let} 
$A_{i} = \overline{\pi}_\Delta(E_i)$,
\emph{for} $1 \leq i \leq e$.
\emph{Let} $A_1, \ldots, A_g$ \emph{be the distinct} $A_{i}$s.
\emph{Then}, \\
$T^+ = \{w | (\exists t_1) \cdots (\exists t_g)(t_1 \in A_{1}^+ \wedge \cdots
\wedge t_g \in A_{g}^+ \wedge w = \{t_1, \ldots, t_g\})\}$ \\
$T_M = \{\langle P,N \rangle | P = \{p_i | (\forall i)p_i \in A_{i}^+\}
\wedge N = \{n_j | (\forall j)n_j \in A_{j}^-\} \wedge i \neq j \wedge |p|
\neq 0 \wedge
 |N| \neq 0 \wedge |P| + |N| = g\}$ \\
$T^- = \{u|(\exists t_1) \cdots (\exists t_g)(t_1 \in A_{1}^- \wedge \cdots 
\wedge t_g \in A_{g}^- \wedge u = \{t_1, \ldots, t_g\})\}$.

\hfill{\space} $\Box$
\end{definition}

%The positive component of the projections consists of the projection of each of the tuple sets onto $\Delta$ and $\overline{\pi}_\Delta(R)^-$ consists of those
%tuple sets in $2^{\tau{(\Delta)}}$, all of whose extensions are in $R^-$.

\begin{definition}
\emph{Let} $R$ \emph{and} $S$ \emph{be normalized extended generalized 
disjunctive 
paraconsistent relations on schemes} $\Sigma$ \emph{and} $\Delta$, 
\emph{respectively with} \\ 
$R_M = \{\langle p_1,n_1 \rangle, \ldots, \langle p_k,n_k \rangle\}$
\emph{and}
$S_M = \{\langle u_1,v_1 \rangle, \ldots, \langle u_m,v_m \rangle\}$.
\emph{Then, the} natural join of $R$ \emph{and} $S$, \emph{denoted} 
$R \dot{\Join} S$, \emph{is a generalized extended disjunctive paraconsistent 
relation on 
scheme} $\Sigma \cup \Delta$, \emph{given by} $R \dot{\Join} S = \gred(T)$,
\emph{where} $T$ \emph{is defined as follows. Let} 
 $E = \{\langle \egred(R)^+ \cup P,\egred(R)^- \cup N
\rangle | P = \{p_i | (\forall i) \langle p_i,n_i \rangle \in \egred(R)_M \}
\wedge N = \{n_j | (\forall j) \langle p_j,n_j \rangle \in$ 
$\egred(R)_M \}
\wedge i \neq j \wedge |P| + |N| = |\egred(R)_M|
 \}$ \emph{and} \\
$F = \{\langle \egred(S)^+ \cup U,\egred(S)^- \cup V \rangle | U = \{u_i | 
(\forall i) \langle u_i,v_i \rangle \in$ \\
$\egred(S)_M \} \wedge V = \{v_j | (\forall j) \langle u_j,v_j \rangle \in 
\egred(S)_M \wedge i \neq j \wedge |U| + |V| = |\egred(S)_M| \}$. \emph{Let the norma
lized elements of} $E$
\emph{be} $E_1, \ldots, E_e$ \emph{and those of} $F$
\emph{be} $F_1, \ldots, F_f$
\emph{and let} $A_{ij} = E_i \overline{\Join} F_j$,
\emph{for} $1 \leq i \leq e$ \emph{and}
$1 \leq j \leq f$.
\emph{Let} $A_1, \ldots, A_g$ \emph{be the distinct} $A_{ij}$s.
\emph{Then}, \\
$T^+ = \{w | (\exists t_1) \cdots (\exists t_g)(t_1 \in A_{1}^+ \wedge \cdots
\wedge t_g \in A_{g}^+ \wedge w = \{t_1, \ldots, t_g\})\}$ \\
$T_M = \{\langle P,N \rangle | P = \{p_i | (\forall i)p_i \in A_{i}^+\}
\wedge N = \{n_j | (\forall j)n_j \in A_{j}^-\} \wedge i \neq j \wedge |p|
\neq 0 \wedge
 |N| \neq 0 \wedge |P| + |N| = g\}$ \\
$T^- = \{u|(\exists t_1) \cdots (\exists t_g)(t_1 \in A_{1}^- \wedge \cdots
\wedge t_g \in A_{g}^- \wedge u = \{t_1, \ldots, t_g\})\}$.

\hfill{\space} $\Box$
\end{definition}

\begin{theorem}
Let $R$ and $S$ be two normalized extended generalized disjunctive 
paraconsistent relations on scheme $\Sigma_1$ and $\Sigma_2$. Also let $F$ be a selection formula on scheme $\Sigma_1$ and  $\Delta \subseteq \Sigma_1$. Then,

\vspace{-0.1in}

\begin{enumerate}
\item $\egrep_{\Sigma_1}(\dot{\sigma}_F(R)) = {\cal S}(\overline{\sigma}_F)
(\egrep_{\Sigma_1}(R))$.
\item $\egrep_{\Sigma_1}(\dot{\pi}_\Delta(R)) = {\cal S}(\overline{\pi}_\Delta)
(\egrep_{\Sigma_1}(R))$.
\item $\egrep_{\Sigma_1 \cup \Sigma_2}(R \dot{\Join} S) = \egrep_{\Sigma_1}(R) 
{\cal S}(\overline{\Join})\egrep_{\Sigma_2}(S)$.
\end{enumerate}
\hfill{\space} $\Box$
\end{theorem}

\section{Conclusions and Future Work}
\label{section5}

We have presented a framework for relational databases under which
disjunctive positive facts, explicit disjunctive negative facts and mixed disjunctive facts can be
represented and manipulated. It is the generalization of generalized 
disjunctive paraconsistent relation in \cite{WAN04}. 
The direction for future work would be to find
applications of the model presented in this paper.
There has been some interest in studying extended disjunctive logic
programs
in which the head of clauses can have one or more literals
\cite{mnk93}. This
leads to two notions of negation: {\em implicit\/} negation 
(corresponding to negative literals in the body) and {\em explicit\/}
negation (corresponding to negative literals in the head).
The model presented in this paper could provide a 
framework under which the semantics of extended
logic programs could
be constructed in a bottom-up manner. 

\vspace{-.1in}

%\begin{thebibliography}{10}

\vspace{-.1in}

\bibliography{/export/home/students/haibin/ismis05/ref}

\end{document}